\documentclass{iaus}
\usepackage{graphicx}

\title[NSSPM, The Prologue] 
{The New Standard Stellar Population Models (NSSPM) -- The Prologue}

\author[H.-c. Lee et al.]   
{Hyun-chul Lee$^1$, Guy Worthey$^1$, Scott C. Trager$^2$, Aaron Dotter$^3$,
Brian Chaboyer$^3$, Jason W. Ferguson$^4$, Darko Jevremovic$^5$, Eddie
Baron$^5$, Paula Coelho$^6$ \and Michael M. Briley$^7$}

\affiliation{$^1$Washington State University \break email: hclee@wsu.edu\\[\affilskip]
$^2$University of Groningen 
$^3$Dartmouth College
$^4$Wichita State University
$^5$University of Oklahoma
$^6$Universidade de Sao Paulo
$^7$University of Wisconsin-Oshkosh}

\pubyear{2007}
\volume{241}  
\pagerange{1--2}
\date{?? and in revised form ??}
\jname{Stellar Populations as Building Blocks of Galaxies}
\editors{Peletier R.F. \& Vazdekis A., eds.}
\begin{document}

\maketitle

\begin{abstract}
We are developing a brand new stellar population models with flexible chemistry  
(isochrones plus stellar colors and spectra) in order to set a new  
standard of completeness and excellence.  Here we present  
preliminary results to assess the effects of stellar evolution models and 
stellar model atmosphere to the well-known Lick indices 
at constant heavy element mass fraction Z that
self-consistently account for varying heavy element mixtures.  We have  
enhanced chemical elements one by one.  Our ultimate goal is to demonstrate 
10 $\%$ absolute mean ages for a sample of local galaxies derived from an 
integrated light spectrum.
\keywords{Stars: abundances, evolution, galaxies: stellar content}
\end{abstract}

\firstsection 
\section{Introduction}
More and more star clusters and galaxies with wide variety of chemical
abundance mixtures are found from detailed high resolution spectroscopic
observations. In order to address and examine those chemically different
stellar systems, we have begun developing a stellar population models with
flexible chemistry.  

\section{Our Preliminary Results}
Dotter et al. (2007, ApJ submitted) have recently calculated stellar 
evolution models with varying chemical abundance mixtures besides 
solar-scaled one. Ten chemical elements (C, N, O, Ne, Mg, Si, S, Ca, Ti, 
and Fe) are individually enhanced by 0.3 dex at a constant total 
metallicity (Z = 0.02). Moreover, additional alpha-enhanced stellar models 
are calculated in order to understand those alpha-element effects collectively 
(for example, Lee \& Worthey 2005). 
Especially, these new sets of stellar evolution models have carefully incorporated 
matching {\em ab initio} low and high temperature opacities.  
Lee et al. (2007, in preparation) will present in full the effects of those
element by element altered isochrones as well as those from synthetic
stellar model atmosphere on top of isochrones effects.

Figure 1 demonstrates the isochrone effects (dots) and that of both
isochrones and stellar model atmosphere (lines) at Fe5270 vs. Mg b diagram. 
It seems that stellar model atmosphere effects are hefty 
for some cases compared to isochrone effects alone.  
Because of our regulation of a fixed total metallicity in this experiment,  
note that oxygen enhancement implies Mg and Fe deficiency, and that 
alpha-enhancement implies mostly depressed Fe abundance.

\begin{figure}
 \includegraphics[width=.99\textwidth]{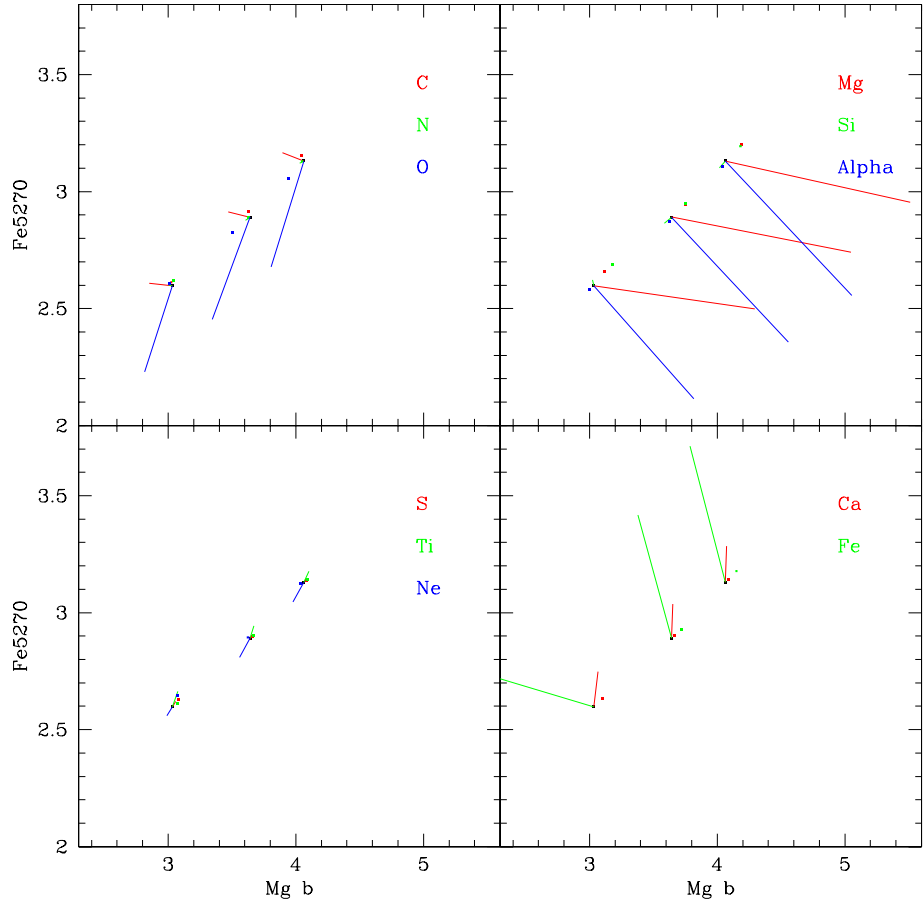}
  \caption{Fe5270 vs. Mg b diagram. At Z = 0.02, individual chemical element 
enhancement effects (plus Alpha-element enhanced case) are shown at 4, 8, 
and 12 Gyr (from left to right in each panels). 
The dots are isochrone effects alone while the lines are isochrone plus stellar model atmosphere effects.}
\end{figure}

\begin{acknowledgments}
Support for this work was provided by the NSF through grant AST-0307487,
the New Standard Stellar Population Models (NSSPM) project. H.-c. Lee also
thanks the organizers for their wonderful hospitality and financial support.
\end{acknowledgments}

\end{document}